# Perspective on Epitaxial NiCo$_2$O$_4$ Film as an Emergent Spintronic Material: Magnetism and Transport Properties

Xiaoshan Xu,[1,a] Corbyn Mellinger,[1] Zhi Gang Cheng,[2,3] Xuegang Chen,[1] and Xia Hong[1,a]

[1] *Department of Physics and Astronomy & Nebraska Center for Materials and Nanoscience, University of Nebraska-Lincoln, Lincoln, Nebraska, 68488-0299, USA*

[2] *Beijing National Laboratory for Condensed Matter Physics and Institute of Physics, Chinese Academy of Sciences, Beijing 100190, China*

[3] *Songshan Lake Materials Laboratory, Dongguan, Guangdong 523808, China*

[a] Authors to whom correspondence should be addressed: xiaoshan.xu@unl.edu and xia.hong@unl.edu

**Abstract**

The ferrimagnetic inverse spinel NiCo$_2$O$_4$ has attracted extensive research interests for its versatile electrochemical properties, robust magnetic order, high conductivity, and fast spin dynamics, as well as its highly tunable nature due to the closely coupled charge, spin, orbital, lattice, and defect effects. Single-crystalline epitaxial thin films of NiCo$_2$O$_4$ present a model system for elucidating the intrinsic physical properties and strong tunability, which are not viable in bulk single crystals. In this perspective, we discuss the recent advances in epitaxial NiCo$_2$O$_4$ thin films, focusing on understanding its unusual magnetic and transport properties in light of crystal structure and electronic structure. The perpendicular magnetic anisotropy in compressively strained NiCo$_2$O$_4$ films is explained by considering the strong spin-lattice coupling, particularly on Co ions. The prominent effect of growth conditions reveals the complex interplay between the crystal structure, cation stoichiometry, valence state, and site occupancy. NiCo$_2$O$_4$ thin films also exhibit various magnetotransport anomalies, including linear magnetoresistance and sign change in anomalous Hall effect, which illustrate the competing effects of band intrinsic Berry phase and impurity scattering. The fundamental understanding of these phenomena will facilitate the functional design of NiCo$_2$O$_4$ thin films for nanoscale spintronic applications.

Table of Contents

## 1. Introduction

Transition metal complex oxides with perovskite ($ABO_3$) and spinel ($AB_2O_4$) structures exhibit diverse functional properties such as magnetism, ferroelectricity, and superconductivity, which are highly tunable due to the close interplay between the charge, spin, orbital, and lattice degrees of freedom.[1] $NiCo_2O_4$ (NCO) belongs to the family of spinel oxides made of magnetic transition metal elements Fe, Co, and/or Ni (see Fig. 1(a) and Table I).[2–8] Each crystal unit cell (uc) consists of 16 octahedral ($O_h$) sites and 8 tetrahedral ($T_d$) sites for the cations (Fig. 1(b)-(c)). The complex chemical/oxygen environment makes these materials highly susceptible to chemical disorder.[9–11] Compared with other members of this family, NCO stands out for its high conductivity ($\sim 10^3\ \Omega^{-1}\mathrm{cm}^{-1}$),[12] potentially half metallicity,[13] fast spin dynamics,[14] and large magnetoelastic effects.[15] The last effect leads to strain tunable magnetic anisotropy, which can result in strong perpendicular magnetic anisotropy (PMA) in epitaxial thin films.[15] In addition, the multi-cation, mixed-valence character makes NCO surface active in charge storage and catalytic processes.[10,11,16–18]

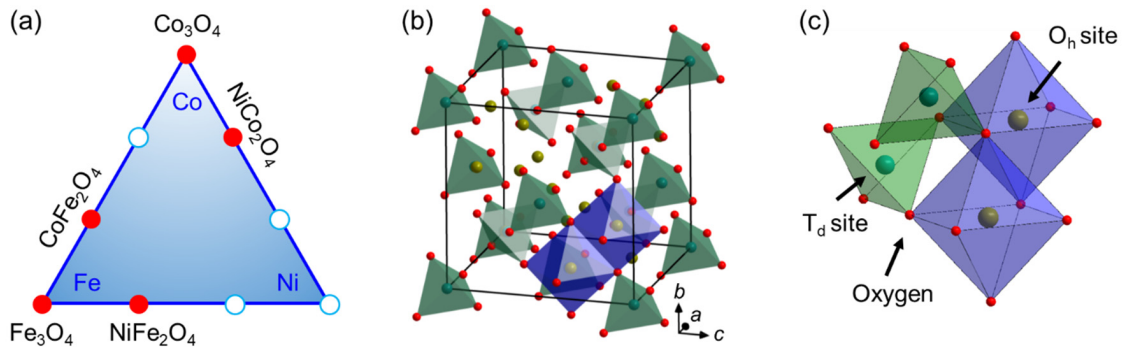

Figure 1. (a) Phase diagram of Fe-Co-Ni based oxide spinel. (b) Crystal structure of spinel oxide consisting of 16 $O_h$ sites (only 2 are shown) and 8 $T_d$ sites per unit cell, with (c) an expanded view illustrating the connection between the neighboring sites.

| Spinel | $Co_3O_4$ Normal[19] | $NiCo_2O_4$ Inverse | $NiFe_2O_4$ Inverse | $CoFe_2O_4$ Inverse | $Fe_3O_4$ Inverse |
|---|---|---|---|---|---|
| $O_h$ site | $Co^{3+}$ | $Ni^{2.5+}$, $Co^{3+}$ | $Ni^{2+}$, $Fe^{3+}$ | $Co^{2+}$, $Fe^{3+}$ | $Fe^{3+}$, $Fe^{2+}$ |
| $T_d$ site | $Co^{2+}$ | $Co^{2.5+}$ | $Fe^{3+}$ | $Fe^{3+}$ | $Fe^{3+}$ |
| Saturation magnetization $M_s$ ($\mu_B$/f.u.) | | 2.2[12] | 2.1[20] | 3.5[21] | 4.0[4] |
| Magnetic order | Antiferromagnetic[19] | Ferrimagnetic[22,23] | Ferrimagnetic[8] | Ferrimagnetic[7] | Ferrimagnetic[7] |
| Magnetic transition temperature | $T_N$ = 40 K[19] | $T_C$ = 420 K[12] | $T_C$ = 840 K[8] | $T_C$ = 790 K[7] | $T_C$ = 860 K[4,7] |
| Room temperature conductivity $\sigma$ (S/cm) | $10^{-4}$ [7] | $10^3$ [10,11,18,24,25] | $10^{-4}$ [6] | $10^{-7}$ [3] | $10^2$ [5] |
| Lattice constant (nm) | 0.8082[26] | 0.8114[27] | 0.834[28] | 0.8381[29] | 0.8396[30] |

Table I. Properties of various Fe, Co, and Ni based spinel oxides.

To date, NCO of spinel structure has been synthesized in the forms of polycrystalline powders and nanocrystals using wet chemical method[10,11,18,24,25,31–33] and, more recently, epitaxial thin films using physical vapor deposition (pulsed laser deposition and RF magnetron sputtering),[9,12,22,32,34–40] while the preparation of large single crystals has not been reported yet. Previous studies of NCO have mainly focused on the electrochemical properties of NCO nanocrystals for catalysis and energy applications.[10,11,41–44] There are emerging interests in single-crystalline epitaxial thin films of NCO, which can not only facilitate the fundamental exploration and nanoscale control of its intrinsic physical properties, but also provide a versatile platform for developing high speed spintronic applications. In this perspective, we discuss the recent advances in the study of epitaxial $NiCo_2O_4$ thin films, focusing on the understanding of their unusual magnetic and transport properties, the impacts of crystal structure and electronic structure, the effects of disorder, strain, and film thickness on property tuning, as well as their potential for spintronic applications.

2. Properties of NCO with Optimal Configuration

It is known that the chemical configurations of NCO, including cation stoichiometry, valency, and site occupancy, depend sensitively on the material preparation conditions.[22,23,35,45–48] In the optimal configuration, NCO possesses a saturation magnetization ($M_S$) of about 2 $\mu_B$/formula unit (f.u.), magnetic Curie temperature ($T_C$) of about 420 K, and conductivity of about $10^3$ $\Omega^{-1}cm^{-1}$.[23,45–47] In this section, we discuss the optimal structure of NCO and compare its properties with other spinel materials (Table I).

2.1 Crystal Structure

As shown in Fig. 1(b), the unit cell of spinel oxide has 8 $T_d$ sites and 16 $O_h$ sites for cations.[49] Regarding the atomic distances, the nearest neighbors are between the edge-sharing $O_h$ sites, and the next nearest neighbors are between the corner-sharing $O_h$ and $T_d$ sites (Fig. (1c)). The third nearest neighbors are between the $T_d$ sites, which are not directed connected. NCO can be viewed as $Co_3O_4$[19,50–52] with one Co replaced by Ni per formula unit. In $Co_3O_4$, the $O_h$ sites are occupied by $Co^{3+}$ and the $T_d$ sites are occupied by $Co^{2+}$, respectively, making $Co_3O_4$ a normal spinel.[53] For NCO, instead of having all the Ni (Co) ions on the $T_d$ ($O_h$) sites, the $T_d$ sites are actually occupied by 8 Co ions, leaving the $O_h$ sites shared by 8 Ni and 8 Co ions.[22,47] This type of arrangement is known as spinel inversion.

## 2.2 Electronic Structure

For NCO, 8 $O_h$ sites in the unit cell are occupied by $Co^{3+}$, and the other 8 $O_h$ sites are occupied by Ni ions with mixed valence ($Ni^{2+}$ and $Ni^{3+}$) (Fig. 2).[9,22] The 3d states of Ni are split into $t_{2g}$ and $e_g$ groups due to the crystal field generated by the surrounding oxygen octahedron, with the $t_{2g}$ states at lower energy (Fig. 2(a)).[54] In comparison, the $T_d$ sites are occupied by the Co ions with mixed valence ($Co^{2+}$ and $Co^{3+}$), with the $e_g$ states at the lower energy (Fig. 2(b)).[54]

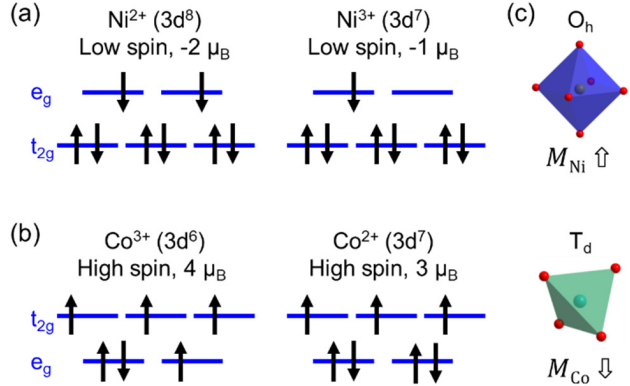

Figure 2. Electronic structure of mixed valence of (a) Ni and (b) Co ions. (c) The spin sub-lattices on the $O_h$ and the $T_d$ sites for NCO in the optimal configuration.

## 2.3 Magnetism

### 2.3.1 Ionic Magnetic Moment

The magnetic properties of NCO hinge on the magnetic moments on the Ni and Co ions, which are related to the site occupancy, valence, and spin states of these ions. The spin state of the ions depends on the local oxygen environment ($O_h$ or $T_d$), as shown in Table II. For the same local environment symmetry, the spin state is determined by the competition between the crystal field splitting energy and the electrostatic energy between two electrons of opposite spins in the same state. When the crystal field splitting energy is small, the electronic configuration follows Hund's rule to maximize the total spin (high spin). Otherwise, the electrons prefer filling the low-lying energy states, corresponding to the low spin states.

|  | **Spin state** | $T_d$ | $O_h$ |
|---|---|---|---|
| $3d^5$ ($Fe^{3+}$) | Low |  | $S=1/2$ |
|  | High | $S=5/2$ | $S=5/2$ |
| $3d^6$ ($Fe^{2+}$, $Co^{3+}$) | Low | $S=1$ | **$S=0^*$** |
|  | High | **$S=2^*$** | $S=2$ |
| $3d^7$ ($Co^{2+}$, $Ni^{3+}$) | Low | $S=3/2$ | **$S=1/2^*$** |
|  | High | **$S=3/2^*$** | $S=3/2$ |
| $3d^8$ ($Ni^{2+}$) | Low/High | **$S=1^*$** | **$S=1^*$** |

Table II. Spin states for various ions, where "*" indicates the spin states in NCO.

For spinel oxides (Table I) of smaller lattice constants, such as $Co_3O_4$ and NCO, the higher crystal-field splitting between the $t_{2g}$ and $e_g$ states on the $O_h$ site favors low spin state (Fig. 2(a)). In contrast, the crystal-field splitting of the $T_d$ sites is typically smaller than that of the $O_h$ sites because of the smaller spatial overlap between the 3d states of the cation and the 2p states of neighboring oxygen ions. The $T_d$ sites thus prefers the high spin states. For example, due to the large crystal-field splitting, $Co^{3+}$ on the $O_h$ sites are at the low spin state carrying zero magnetic moment. In $Co_3O_4$, all $O_h$ sites are occupied by the non-magnetic $Co^{3+}$, making $Co^{2+}$ on the $T_d$ sites the only magnetic ions[53]. This explains the low magnetic transition temperature (40 K) of $Co_3O_4$, since the $T_d$ sites are far away from each other (Fig. 1(c)).[19] On the other hand, for those of larger lattice constants in Table I ($NiFe_2O_4$, $CoFe_2O_4$, and $Fe_3O_4$), the larger octahedron and tetrahedron lead to smaller crystal-field splitting, making the high-spin states more energetically favorable [4,20,21,55]. The sensitive dependence on lattice constants also means the magnetic states of spinel materials can be effectively altered using epitaxial strain.[56]

As shown in Fig. 2, for NCO, the magnetic moments are carried by $Ni^{2+}$ and $Ni^{3+}$ on the $O_h$ sites and $Co^{2+}$ and $Co^{3+}$ on the $T_d$ sites. Here $Ni^{2+}$ and $Ni^{3+}$ are in their low-spin states, giving rise to an average magnetic moment of 1.5 $\mu_B$/Ni. $Co^{2+}$ and $Co^{3+}$ are in their high-spin states, corresponding to an average magnetic moment of 3.5 $\mu_B$/Co.

### 2.3.2 Magnetic Order and Exchange Interaction

The strongest exchange interaction in magnetic spinels is between the corner-sharing $T_d$ and $O_h$ sites.[57] The approximately 125° cation-oxygen-cation bond angle leads to antiferromagnetic interaction.[58,59] The second largest exchange interaction is between the edge-sharing $O_h$ sites, where the 90° cation-oxygen-cation bond angle corresponds to ferromagnetic interaction, which is weaker than the antiferromagnetic interactions in general [58,59]. The mixed valence on the $O_h$ sites also enables the potential double exchange interaction, which is related to the metallicity of NCO.[22]

Since the magnetic Ni and Co ions occupy the corner-sharing $O_h$ and $T_d$ sites, respectively, their magnetic moments are expected to be antiparallel (Fig. 2(c)). The cancellation between the two spin sub-lattices is incomplete due to the different magnetic moments on Ni and Co ions, resulting in ferrimagnetism (Fig. 3(a)).[23] Given that the average spin magnetic moments are 1.5 $\mu_B$/Ni on $O_h$ sites and 3.5 $\mu_B$/Co on $T_d$ sites, the optimal magnetic moment per formula unit is 2.0 $\mu_B$, which is consistent with the experimental observation of high quality epitaxial thin films (Fig. 3(b)).[12,23]

Figure 3(b) shows the magnetic properties of (001) NCO films deposited on $Mg_2AlO_4$ (MAO) substrates. The magnetic Curie temperature is about 420 K for the 30 uc film, as deduced by extrapolating $M(T)$ beyond 400 K (Fig. 3(b)).[12] It is lower than many other spinels listed in Table I, such as $Fe_3O_4$, $NiFe_2O_4$, and $CoFe_2O_4$.[4,7] According to the mean field theory, this is likely related to the coordination number for the magnetic ions. Considering only the strongest $T_d$-$O_h$ exchange interactions, each $T_d$ site has 12 neighboring $O_h$ sites. When 50% of the $O_h$ sites are occupied by the nonmagnetic $Co^{3+}$, the magnetic coordination number is reduced by 50%, meaning the pair of strongest exchange interaction per unit cell is reduced by 50%. This is consistent with the roughly 50% reduction in $T_C$ compared with that of other compounds in Table I except $Co_3O_4$.

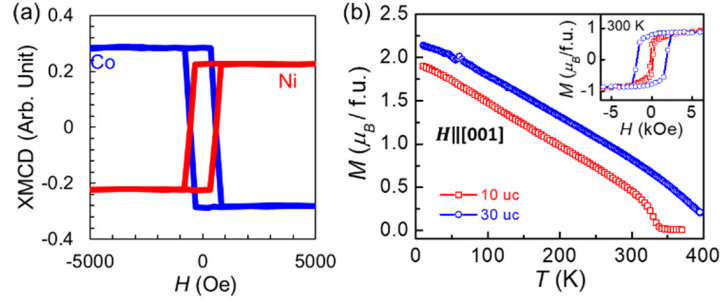

Figure 3. Magnetic properties of optimal NCO(001)/MAO(001) films. (a) X-ray magnetic circular dichroism (XMCD) signal vs. magnetic field taken on a 30 nm film showing antiparallel alignment between the moments on Co and Ni. Adapted from Ref. [23]. Copyright 2020 American Physical Society. (b) $M$ vs. $T$ and $M$ vs. $H$ (inset) taken on 10 and 30 uc films showing low temperature $M_S$ of about 2.2 $\mu_B$/f.u. and $T_C$ of about 420 K from extrapolation. Adapted from Ref. [12]. Copyright 2019 John Wiley & Sons, Inc.

### 2.3.3 Magnetic Anisotropy

Besides being ferrimagnetic above room temperature, one appealing property of NCO is its high magnetic anisotropy energy (MAE), which can be sensitively tuned by epitaxial strain. Since NCO has a cubic structure (lattice constant of 0.8114 nm[27]), using the minimum number of terms, the free energy of magnetic anisotropy can be described by:[60]

$$F(\hat{m}) = K_1(m_x^2 m_y^2 + m_y^2 m_z^2 + m_z^2 m_x^2), \qquad (1)$$

where $\hat{m} = (m_x, m_y, m_z)$ is the unit magnetization vector and $m_x, m_y, m_z$ are the projection of $\hat{m}$ on the $\hat{x}$, $\hat{y}$, and $\hat{z}$ axes ($\Sigma m_i^2 = 1$), respectively. The coefficient $K_1$ measures the MAE. If $K_1 > 0$, the easy axes are along the {100} directions. Otherwise, the easy axes are along the {111} directions.

As bulk single crystals of NCO are not available, single-crystalline NCO can only be synthesized in the form of epitaxial films, which are slightly deformed from the cubic symmetry by the epitaxial strain. In these systems, the MAE can be determined by measuring epitaxial thin films grown along different crystalline orientations. For a strained epitaxial film, the free energy can be modified to include the magnetoelastic effects:[60]

$$F_{ME}(\hat{m}) = K_1(m_x^2 m_y^2 + m_y^2 m_z^2 + m_z^2 m_x^2) + B_1(m_x^2 e_{xx} + m_y^2 e_{yy} + m_z^2 e_{zz}) + B_2(m_x m_y e_{xy} + m_y m_z e_{yz} + m_z m_x e_{zx}), \qquad (2)$$

where $e_{ij}$ ($i,j=x, y, z$) are the components of the strain tensor, and $B_1$ and $B_2$ are the longitudinal and shear magnetoelastic coupling constants, respectively. Figure 4 shows schematically the non-zero strain components $e_{ij}$ for NCO films grown on substrates of different orientations.

For the (001) oriented substrates, the non-zero components of the strain tensor are $e_{zz}$, and $e_{xx} = e_{yy}$. The free energy of the in-plane and out-of-plane directions are $F_{ME}(\hat{x}) = B_1 e_{xx}$ and $F_{ME}(\hat{z}) = B_1 e_{zz}$, respectively.

For the (110) oriented substrates, the non-zero components of the strain tensor are $e_{xx} = e_{yy}$, $e_{zz}$, and $e_{xy}$. The free energy of the out-of-plane direction is $F_{ME}(\hat{m}_{(110)}) = \frac{K_1}{4} + B_1 e_{xx} + \frac{B_2}{2} e_{xy}$. For the in-plane [001] direction, $F_{ME}(\hat{m}_{(001)}) = B_1 e_{zz}$.

For the (111) oriented substrates, the non-zero components of the strain tensor are $e_{xy} = e_{yz} = e_{zx}$. The free energy of the out-of-plane direction is $F_{ME}(\hat{m}_{(111)}) = \frac{K_1}{3} + B_2 e_{xy}$. For the two in-plane directions, one has $F_{ME}(\hat{m}_{(\bar{1}10)}) = \frac{K_1}{4} - \frac{B_2}{2} e_{xy}$, and $F_{ME}(\hat{m}_{(1\bar{1}2)}) = \frac{K_1}{4} - \frac{B_2}{2} e_{xy}$.

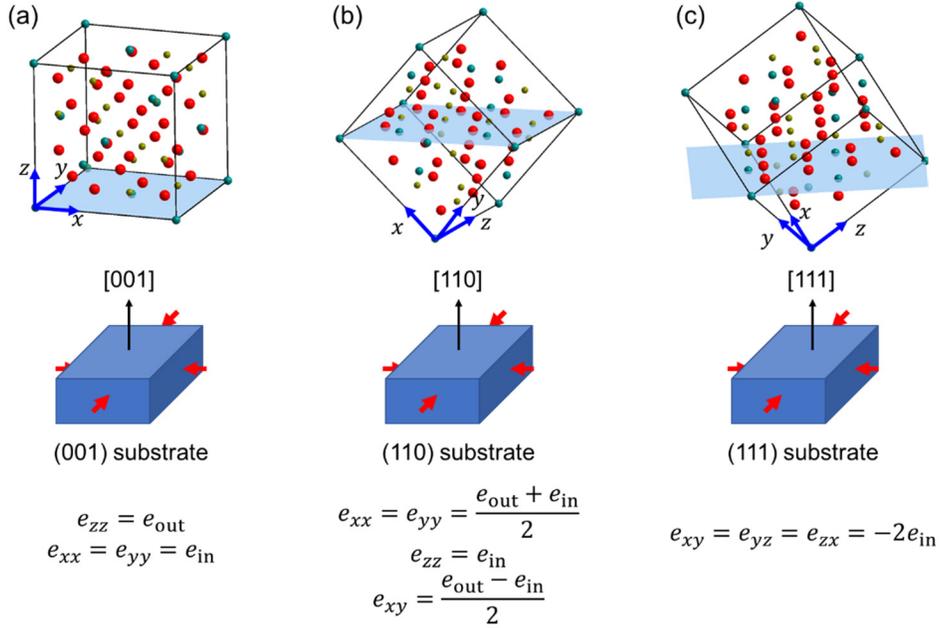

Figure 4. Illustration of epitaxial NCO films grown along the (001) (a), (110) (b), and (111) (c) orientations respectively, and the corresponding non-zero epitaxial strain $e_{ij}$ (i, j=x, y, z). Here $e_{in}$ and $e_{out}$ are strain along the in-plane and out-of-plane directions, respectively.

Experimentally, when the NCO films are grown on MAO substrates (lattice constant of 8.09 Å),[61,62] the substrates exert an in-plane compressive strain $e_{in}$= -0.3% on the film. For the NCO (001) films, the strain can be measured as $e_{xx} = e_{in} < 0$ and $e_{zz} = e_{out} > 0$, where $e_{out}$ reflects the difference between the spacing of the (001) plane for the strained film and that of the bulk values. As shown in Fig. 5(a), the easy axis is along the out-of-plane [001] direction, while the in-plane [100] direction becomes a hard axis. Defining the energy integral $I = \int_0^{H_{sat}} M dH$, where $H$, $H_{sat}$, and $M$ are the magnetic field, saturation magnetic field for the hard axis, and magnetization, respetively, the free energy different between the in-plane [100] direction and out-of-plane [001] direction can be calculated as $E_a = I_{[100]} - I_{[001]}$, which is equal to $F_{ME}(\hat{x}) - F_{ME}(\hat{z}) = B_1(e_{xx} - e_{zz})$. This mean that $B_1 = \frac{E_a}{e_{in} - e_{out}}$.

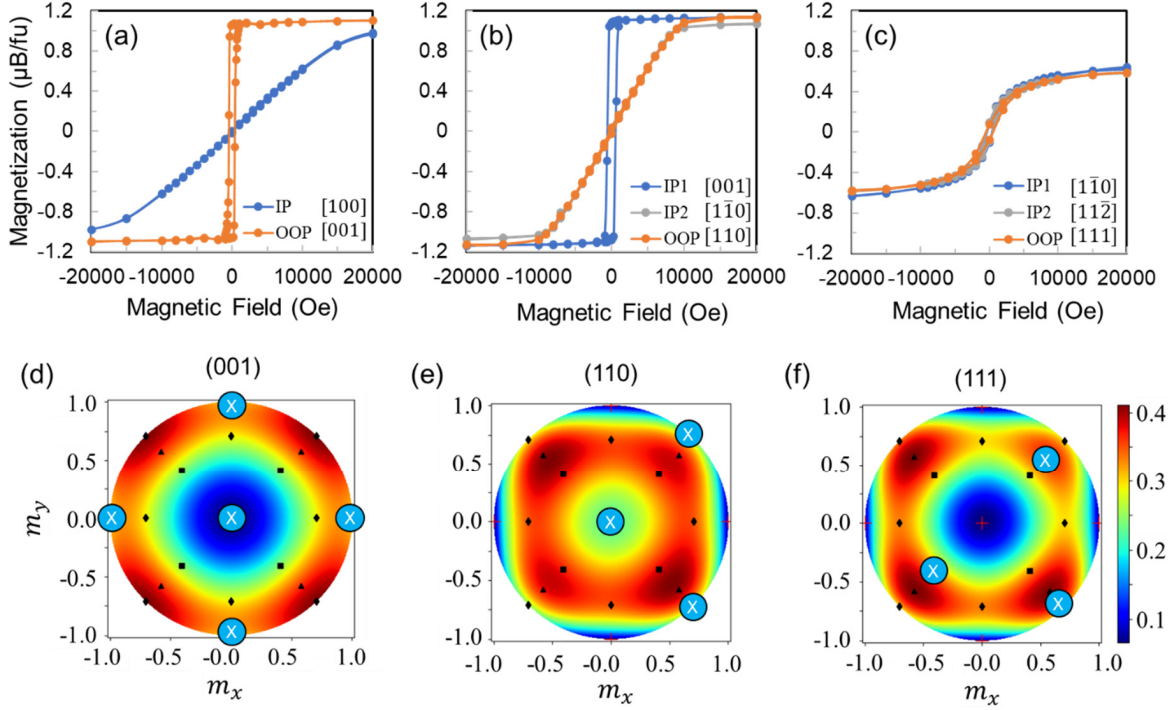

Figure 5. Magnetic hysteresis measured along in-plane and the out-of-plane directions for (a) NCO(001)/MAO(001), (b) NCO(110)/MAO(110), and (c) NCO(111)/MAO(111) films (about 15 nm) at 150 K. (d), (e), (f) are the simulated free energy for (001), (110), (111) films, respectively. The "×" symbols mark the directions along which the magnetic hysteresis is measured. Adapted from Ref. [15]. Copyright 2020 American Physical Society.

For NCO (110) films, the strain can be measured as $e_{xx} = e_{yy} = \frac{e_{out}+e_{in}}{2} > 0$, $e_{xy} = \frac{e_{out}-e_{in}}{2} > 0$, and $e_{zz} = e_{in} < 0$. As shown in Fig. 5(b), both the in-plane $[1\bar{1}0]$ and out-of-plane [110] directions are hard axes. Hence, the free energy difference between the [110] and [001] directions is $E_a = I_{[110]} - I_{[001]}$, which is equal to $F_{ME}(\hat{m}_{(110)}) - F_{ME}(\hat{z}) = \frac{K_1}{4} + (2B_1 + B_2)\frac{e_{out}-e_{in}}{4}$. This can be measured using the hysteresis loops in Fig. 5(b). Considering $e_{out} - e_{in} \ll 1$, $E_a = \frac{K_1}{4} + (2B_1 + B_2)\frac{e_{out}-e_{in}}{4} \approx \frac{K_1}{4}$.

For the NCO (111) films, the strain can be measured as $e_{xy} = e_{yz} = e_{zx} = -2e_{in} < 0$. As shown in Fig. 5(c), the hysteresis loops taken along the in-plane $[1\bar{1}0]$ and $[11\bar{2}]$ directions as well as the out-of-plane [111] direction are similar. By equating the free energy, one has $K_1 = -18B_2 e_{xy} = -36B_2 e_{out}$.

By measuring $e_{out}$ and $E_a$ of the (001) and (110) films at 20 K, one finds $K_1 \approx 0.76$ MJ/m³, $B_1 \approx$ -28 MJ/m³, and $B_2 \approx$ -7.0 MJ/m³. These values are comparable to those of CoFe$_2$O$_4$, which is well-known for the high MAE and the magnetoelastic effects.[60] The room temperature values of these constants are about 40% of the low temperature values, as determined by the reduced magnetization (Fig. 3(b)).

Figures 5(d)-(f) display the simulated free energy using the extracted $K_1$, $B_1$, and $B_2$ values and the measured strain. As shown in Fig. 5(d), as $B_1 < 0$, the compressive strain increases the energy along the [100] and [010] directions, making them the hard axes. Meanwhile, the energy along the [001] direction is reduced, making it a global easy axis. Hence, NCO(001)/MAO(001) possesses uniaxial PMA. For the NCO(110) films, the strain mainly makes the [001] direction a local easy axis. For the NCO(111) films, the [100], [010], and [001] remain easy axis, while the energy of the [111] axis is reduced to be comparable to that of the [$\bar{1}$10] axis.

3. Tuning of Magnetic Properties

This section discusses the tuning of the magnetic properties in NCO thin films by the structural and chemical variation from the optimal configurations.

3.1 Spin-Lattice Coupling and Perpendicular Magnetic Anisotropy in Epitaxial Thin Films

The large magnetoelastic constant suggests that the magnetic properties of NCO are highly tunable using mechanical means via spin-lattice coupling. Conversely, it is also possible to change the shape of NCO slightly using the magnetic field. In particular, NCO(001) films strained on MAO exhibits a strong PMA, with $K_u$ = 0.45 MJ/m$^3$ at 20 K and 0.21 MJ/m$^3$ at 300 K[47,63] (Table III).

| Material/Structure | $K_u$ (MJ/m$^3$) |
|---|---|
| La$_{0.7}$Sr$_{0.3}$MnO$_3$ single crystal | 0.0018 (300 K) [64] |
| NiCo$_2$O$_4$ (001)/MgAl$_2$O$_4$ (001) | 0.45 (20 K)<br>0.21 (300K) [47,63] |
| CoFe$_2$O$_4$ (001) / MgO(001) | 0.22 [65]<br>0.97 [66] |
| CoFeB/MgO (001) | 0.21 [67] |
| Co/Ni multilayer | 0.5 [68] |
| Co/Pt multilayer | 0.9 [69] |
| CoPt single crystal | 4.1 [70] |

Table III. Comparison of MAE $K_u$ in various magnetic material systems.

The ability to achieve high magnetic anisotropy has significant impact on developing energy and information applications. For example, electrodes with strong PMA are highly desired for nanoscale spintronic devices to achieve high thermal stability and energy-efficient switching.[71] Most materials or heterostructures of high PMA are based on intermetallic compounds,[72–76] multilayers,[68,69] or metal/oxide interfaces,[71] with many of them involving high-cost elements such as Au and Pt. In contrast, transition-metal oxide conductors, despite their advantages of low-cost as well as structural and chemical stabilities, have rarely been reported to exhibit high PMA (Table III).

Magnetic anisotropy originates from structural anisotropy and spin-orbit coupling (SOC). In ordered intermetallic compounds containing strong SOC nonmagnetic (NM) metals (*e.g.*, Pd, Au, and Pt) and ferromagnetic (FM) metals (*e.g.*, Fe and Co), anisotropic crystal structures lead to anisotropic hybridization between the states in the NM and FM elements and consequently high magnetic anisotropy (about 5 MJ/m$^3$).[72–77] The structural anisotropy can also be introduced by stacking NM and FM layers for high PMA (about 1 MJ/m$^3$).[69] On the other hand, remarkable PMA has been demonstrated in Co/Ni multilayers (about 0.5 MJ/m$^3$)[68] and FM/oxide interfaces (about 0.2 MJ/m$^3$)[71,78] without the need of strong SOC NM. Here, the electronic degeneracy and occupancy are adjusted such that the 3d states in FM with a large orbital angular momentum (in-plane states) determine the magnetic anisotropy. In particular, at the FM/oxide interface, the 3d electronic states are tuned via the hybridization with oxygen states. This suggests the possibility of having transition-metal oxides with high magnetic anisotropy.

In 3d transition-metal oxides, the hybridization of the metal 3d and oxygen 2p states generates a crystal-field splitting of $\Delta \sim 1$ eV; the SOC ($\xi \sim 50$ meV) couples these split states and modifies the energy by $\approx \langle L_z \rangle \xi^2/\Delta$, where $\langle L_z \rangle$ is the average angular momentum projected along the out-of-plane direction. This energy modification, which gives rises to the MAE, can reach $\sim 1$ meV for $\langle L_z \rangle \sim 1$. This is the origin of the large magnetic anisotropy ($\sim 1$ MJ/m$^3$) and strong spin-lattice coupling in CoFe$_2$O$_4$, an insulator that exhibits PMA in strained films.[66,78–88] For oxide conductors, however, room temperature ferromagnetism is already rare, not to mention high magnetic anisotropy. Most widely studied FM oxide conductors, such as (La,Sr)MnO$_3$, have low magnetic anisotropy[89,90] and weak spin lattice coupling due to the dominant $z^2$ state, which has a low orbital angular momentum.[91]

The microscopic origin of the large magnetoelastic constants has to do with the electronic structure of NCO and particularly the effect of strain on the single-ion MAE via SOC. A model Hamiltonian can be written using a one-electron picture:

$$H = H_{\text{CF}} + \frac{\xi}{\hbar^2}\vec{s}\cdot\vec{l} + \frac{E_{ex}}{\hbar}\vec{s}\cdot\hat{A}. \qquad (3)$$

Here $H_{\text{CF}}$ is the Hamiltonian including the crystal field, $\vec{l}$ and $\vec{s}$ are the orbital and spin angular momentum, respectively, and $E_{\text{ex}}$ and $\hat{A}$ are parameters for simulating the exchange interaction between the ion with the neighboring magnetic ions with spins along the direction $\hat{A}$. The magnetoelastic coupling can be understood as follows: when strain modifies the electronic orbital states by changing the local environment of the magnetic ions ($H_{\text{CF}}$), the preferred spin orientation also changes due to the SOC.

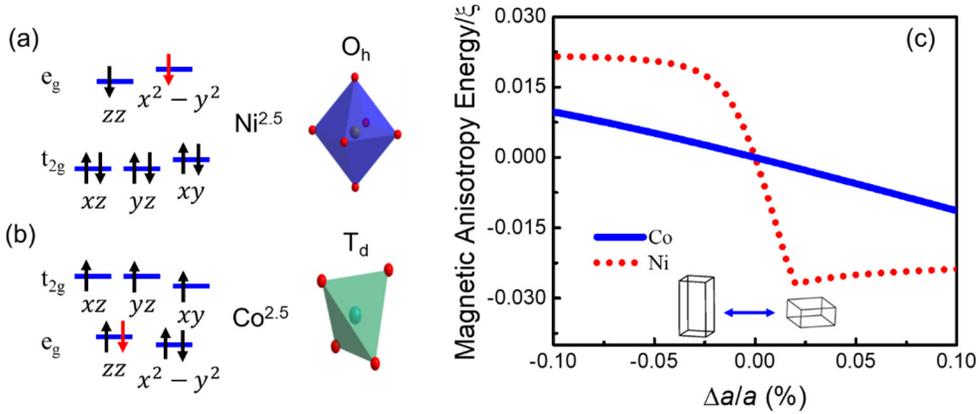

Figure 6. Microscopic model of the effect of biaxial strain in (001) plane. (a)-(b) Energy diagram of Ni and Co ions on the O$_h$ and T$_d$ sites, respectively, showing the splitting between e$_g$ and t$_{2g}$ levels, as well as the smaller splitting within the symmetry groups due to the tetragonal distortions under compressive strain. The red arrows represent partial occupation of the orbital. (c) MAE calculated from the single-ion magnetic anisotropy as a function of in-plane biaxial strain, where $\xi$ = 50 meV is the SOC constant. The magnitude of $H_{\text{CF}}$ splitting and $E_{\text{ex}}$ are set as 1 eV and 5 eV, respectively. Adapted from Ref. [15]. Copyright 2020 American Physical Society.

In one unit cell of NCO, there are 8 low-spin $Ni^{2.5}$ ions in the $NiO_6$ octahedron (Figs. 6(a), $O_h$ symmetry) and 8 high-spin $Co^{2.5}$ ions in the $CoO_4$ tetrahedron (Fig. 6(b), $T_d$ symmetry), where the fractional valence reflects the mixed valence state.[22,34] The Co and Ni 3d states are split into doubly degenerate $e_g$ states and triply degenerate $t_{2g}$ states due to the corresponding $H_{CF}$. Under the biaxial compressive strain, which reduces the cubic symmetry to tetragonal, these states further split (Fig. 6(a)-(b)).

Figure 6(c) shows the simulated single-ion magnetic anisotropy $E_{SIMA}$ as a function of strain in (001) NCO films, assuming $\xi$, crystal field splitting, and $E_{ex}$ as 0.05, 1, and 5 eV, respectively. The crystal field is simulated by replacing the oxygen atoms with point charges in $NiO_6$ and $CoO_4$. The total energy on a magnetic ion $E_t$ is calculated by summing the energy of the individual electrons[68] according to the population of d-orbitals (Figs. 7(a)-(b)). The single-ion magnetic anisotropy is manifested in the dependence of $E_t$ on the direction of $\hat{A}$, defined as $E_{SIMA} = E_{t,x} - E_{t,z}$ for (001) NCO, where $E_{t,x}$ and $E_{t,z}$ are the total energy for $\hat{A} = \hat{x}$ (in-plane) and $\hat{A} = \hat{z}$ (out-of-plane), respectively. The epitaxial strain $\Delta a/a$, where $a$ is the bulk lattice constant, is introduced by distorting the $NiO_6$ and $CoO_4$ local environment according to the lattice constant change, which are $\Delta a$ and $-2\Delta a$ for in-plane and out-of-plane axes, respectively. For both $Ni^{2.5}$ and $Co^{2.5}$, under the tetragonal distortion due to the compressive strain ($\Delta a < 0$), $E_{SIMA}$ is positive. This means that the $c$ axis (out-of-plane direction) is the easy axis, which is consistent with the experimental observation.[12,15,37]

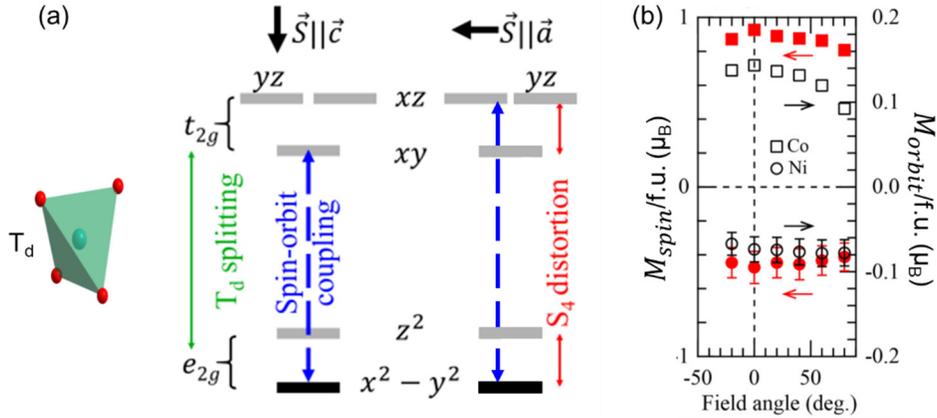

Figure 7. (a) Relative energies of mixed states due to the SOC. The energy gain is larger when the spin is along the $c$ axis than that when the spin is perpendicular to the $c$ axis, leading to magnetocrystalline anisotropy. Adapted from Ref. [15]. Copyright 2020 American Physical Society. (b) Measured spin and orbital contributions to the magnetization for Co and Ni in NCO(001)/MAO(001) (about 30 nm). The orbital moment of Co is not only more substantial than that of Ni, but also shows a clear anisotropy toward the out-of-plane direction. Adapted from Ref. [23]. Copyright 2020 American Physical Society.

A microscopic understanding of the effect of strain on magnetic anisotropy can be gained using the $Co^{2.5}O_4$ tetrahedron as an example (Fig. 6(c)). In this case, the tetragonal distortion generates an $S_4$ symmetry (Fig. 6(b)). The 3d electronic configuration can thus be viewed as a half-filled shell plus an electron in the $|x^2-y^2\rangle$ state and a fractional occupation of the $z^2$ state. Since the half-filled shell is not expected to contribute to the magnetic anisotropy, the electron in the $|x^2-y^2\rangle$ state dominates the anisotropy energy. As illustrated in Fig. 7(a), if the spin is along the $z$ axis, the $|x^2-y^2, s_z=½\rangle$ state couples to the $|xy,$ $s_z=½>$ state to lower its energy, with a coupling strength $<x^2-y^2, s_z=½|\frac{\xi}{\hbar^2}\vec{s}_i \cdot \vec{l}_i| xy, s_z=½> = \xi$. On the other hand, when the spin is along the $x$ axis, the $|x^2-y^2, s_x=½>$ state couples to the $|xz, s_x=½>$ state to lower its energy, with a coupling strength $<x^2-y^2, s_x=½|\frac{\xi}{\hbar^2}\vec{s}_i \cdot \vec{l}_i| xz, s_x=½> = \xi/2$, which is smaller than that when the spin is along the $c$ axis. Therefore, the compressive strain results in an out-of-plane magnetic anisotropy. Hence, the 3d $|x^2-y^2>$ state of Co in the Co$^{2+\delta}$O$_4$ tetrahedron plays a key role in the magnetoelastic coupling of NCO due to its potentially large orbital angular momentum along the $z$ axis. Assuming the magnitude of $\xi$, crystal field splitting, and $E_{ex}$ as 0.05, 1 and 5 eV, respectively, the single-ion magnetic anisotropy is found to be about 1 meV/f.u. (Fig. 6(c)). This translates to about 1 MJ/m$^3$ in MAE, in good agreement with the observed values in Table III. Recent studies further show that the tetragonal distortion is responsible for the spin reorientation transition from PMA to an easy cone anisotropy in NCO at low temperature.[92] The dominant contribution of Co to the PMA has been confirmed by a recent study using XMCD, which reveals orbital magnetic moments of 0.14 μ$_B$/Co and 0.07 μ$_B$/Ni (Fig. 7(b)).[23] In addition, the orbital magnetic moment of Co is along the out-of-plan direction, while that of Ni is more isotropic.[23]

### 3.2 Effects of Cation Stoichiometry, Valence, and Site Occupancy

Ever since the first stabilization of NCO in the form of epitaxial thin films, the high tunability of its physical properties by the growth condition has gained attention[9,22,35–40]. For example, NCO can be tuned from a conductor to an insulator by changing the growth temperature;[22,35] the magnetic transition temperature and saturation magnetization can be varied substantially using different O$_2$ pressure during film growth.[47] Chemical disorders such as cation/oxygen vacancies, valence disorder, and cation inversion can have significant impacts on the electronic and magnetic properties of NCO.[93] In this section, we discuss the effects of cation stoichiometry, valence, and site occupancy on the observed property variation.

NCO can be viewed as Co$_3$O$_4$ with one Co atom replaced by Ni atom per formula unit. For Co$_3$O$_4$, Co$^{3+}$ occupies the O$_h$ site, which has a higher coordination number. As Co$^{3+}$ is less stable than Co$^{2+}$, such a configuration enhances its bonding and stability. For NCO, the site occupancy is complicated by the additional species of cations (Ni$^{2+}$ and Ni$^{3+}$) and their chemical stability. In principle, Ni may replace up to all the Co$^{2+}$ on the T$_d$ sites and up to 50% of the Co$^{3+}$ on the O$_h$ sites. The large number of possible configurations of NCO involving 4 types of cations (Ni$^{2+}$, Ni$^{3+}$, Co$^{2+}$, and Co$^{3+}$) and two sites (O$_h$ and T$_d$) and the subtle differences in cation stability all depend sensitively on the growth condition, which provides an effective means to tune the properties of NCO.

To describe the stoichiometry and valency of NCO, we use the chemical formula $[Ni_{n1}^{2+}Ni_{n2}^{3+}Co_{n3}^{2+}Co_{n4}^{3+}]Co^{3+}O_4$, where $\Sigma n_i \leq 2$. Here the Co$^{3+}$ outside the bracket occupies the O$_h$ sites. The site occupancy can be described using the fractions of these cations on the O$_h$ sites $x_i$, where $x_i \leq 1$. The additional restriction of charge neutrality requires $2n_1 + 3n_2 + 2n_3 + 3n_4 = 5 - 2V_o$, where $V_o$ is the number of oxygen vacancy per formula unit.

Following the rules of spin states in Table II, one can calculate the total spin magnetization per formula unit in terms of the structure configuration parameters:

$\frac{M_S}{\mu_B} = 2n_1 + 3n_2 + 3n_3 + 4n_4 - 4(n_1x_1 + n_2x_2 + n_3x_3 + n_4x_4) = 5 - 2V_o + n_3 + n_4 - 4(n_1x_1 + n_2x_2 + n_3x_3 + n_4x_4)$. (4)

Note that $n_1x_1 + n_2x_2 + n_3x_3 + n_4x_4$ is the occupancy of the $O_h$ sites combining all cations. Therefore, if the number of each ionic specie ($n_i$) and the occupancy of the $O_h$ sites are fixed, the individual $x_i$ does not matter. In other words, swapping the cations between the $O_h$ and $T_d$ sites (degree of spinel inversion) would not change magnetization. This can be inferred in Table II: for all four cations, moving a cation from an $O_h$ sites to a $T_d$ site increases the magnetic moment by 2 $\mu_B$. Swapping two cations between the $O_h$ and $T_d$ sites would thus increase magnetization for one cation and reduce the contribution of the other by the same amount, with the net magnetization unchanged.

### 3.2.1 Effect of Valency

The effect of valency can be examined by changing the oxygen stoichiometry of NCO, which has been realized by growing NCO films under optimal condition and carrying out post-growth annealing at the growth temperature to generate oxygen vacancies.[45] As shown in Fig. 8(a)-(b), the valency of cations in NCO was characterized using x-ray absorption spectroscopy.[45] After annealing in vacuum at 315 °C, the valence of Co remains mostly unchanged, while for Ni an obvious increase of $Ni^{2+}$ are observed.

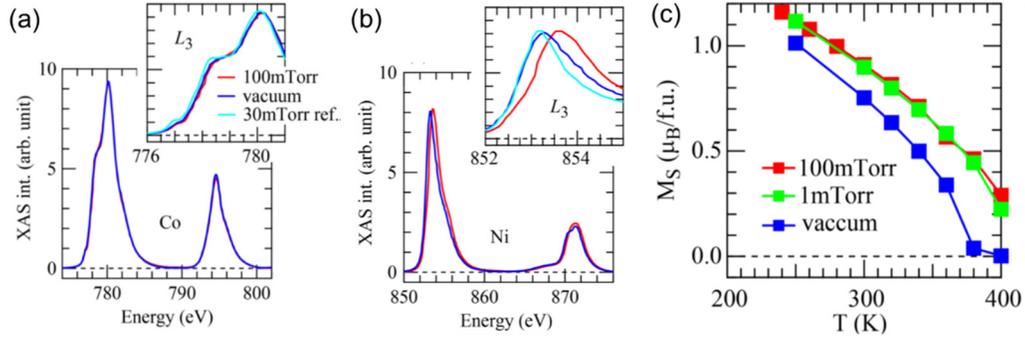

Figure 8. Characterization of NCO(001)/MAO(001) films (about 30 nm) grown in optimal condition with and without post-growth annealing. (a)-(b) X-ray absorption spectra taken on (a) Co and (b) Ni L edges. (c) Temperature dependence of magnetization for films with different annealing conditions. Adapted from Ref. [45]. Copyright 2020 American Institute of Physics.

Starting from nearly optimal structural configuration, annealing in vacuum does not change the total site occupation, or $n_1x_1 + n_2x_2 + n_3x_3 + n_4x_4 = 1$ in Eq. (4). Since Co cations are not affected, $n_3$ and $n_4$ are constants. Therefore, Eq. (2) becomes:

$$\frac{M_S}{\mu_B} = 1 - 2V_o + n_3 + n_4. \tag{5}$$

Clearly, when $V_o$ increases, $M_S$ decreases. This can be understood considering that the contribution of $Ni^{3+}$ to magnetization is larger than that of the $Ni^{2+}$, according to Table II. This is consistent with the observation in Fig. 8(c), where $M_S$ decreases after post-growth annealing in vacuum.[45] In addition, $T_C$ also reduces slightly, reflecting the change of exchange interaction when $Ni^{3+}$ is converted to $Ni^{2+}$.[45]

### 3.2.2 Effect of Cation Stoichiometry

Change of cation stoichiometry of NCO has been observed in epitaxial thin films grown in different oxygen pressure [Fig. 9].[47] As shown in Fig. 9(a)-(b), when the $O_2$ growth pressure decreases, the occupancy of Ni cation on the $O_h$ site decreases substantially. In contrast, the small occupancy of Ni cation on the $T_d$ site remains approximately constant. Overall, when the $O_2$ pressure decreases, total population of the Ni cation decreases. This is consistent with the observation in Fig. 8: $Ni^{3+}$ originally occupies the $O_h$ site in the optimal NCO structure and becomes unstable when oxygen stoichiometry is changed.[45] The change of oxygen vacancy due to the reduced $O_2$ pressure during film deposition further weakens the bonding of Ni cations with the surrounding oxygen, which leads to the reduction of Ni cation population.

Another important observation for NCO films grown in reduced $O_2$ pressure is the cation vacancy on the $T_d$ sites [Fig. 10(a)],[40,48] which can even result in the formation of rock salt (Ni,Co)O phase.[94] The $T_d$ sites are mainly occupied by Co cations in the optimal NCO structure. First-principles calculations indicate that the energy of the $Co^{3+}$ on the $T_d$ site is higher than that of $Ni^{2+}$ on the $O_h$ site, while the latter is slightly higher than that of $Co^{3+}$ on the $O_h$ sites.[40] Given that no clear $O_h$ vacancy is observed during the growth of NCO in reduced $O_2$ pressure, it is likely that the $Co^{3+}$, which occupies the $T_d$ site in the optimal NCO structure, tends to occupy the $O_h$ site instead when the fractional occupancy of Ni is reduced on the $O_h$ site (Fig. 10(b)).

According to Eq. (5), when the total occupancy on the $O_h$ site is fixed, the cation vacancy is expected to significantly reduce $M_s$, which is consistent with the observation in Fig. 9(c).[47] Since the Co cation on the $T_d$ site is mostly responsible for the magnetic anisotropy (Fig. 7), the reduction of Co occupancy on the $T_d$ site is expected to decrease the MAE, which is indeed observed experimentally (Fig. 9(d)).[47] Furthermore, when there is cation vacancy on the $T_d$ site, the effective coordination number for the exchange interaction is reduced, which lowers $T_C$[47] according to the mean field theory. As shown in Fig. 9(c), the change of $T_C$ follows that of $M_s$ closely, suggesting the key role played by the cation vacancy on the $T_d$ site.

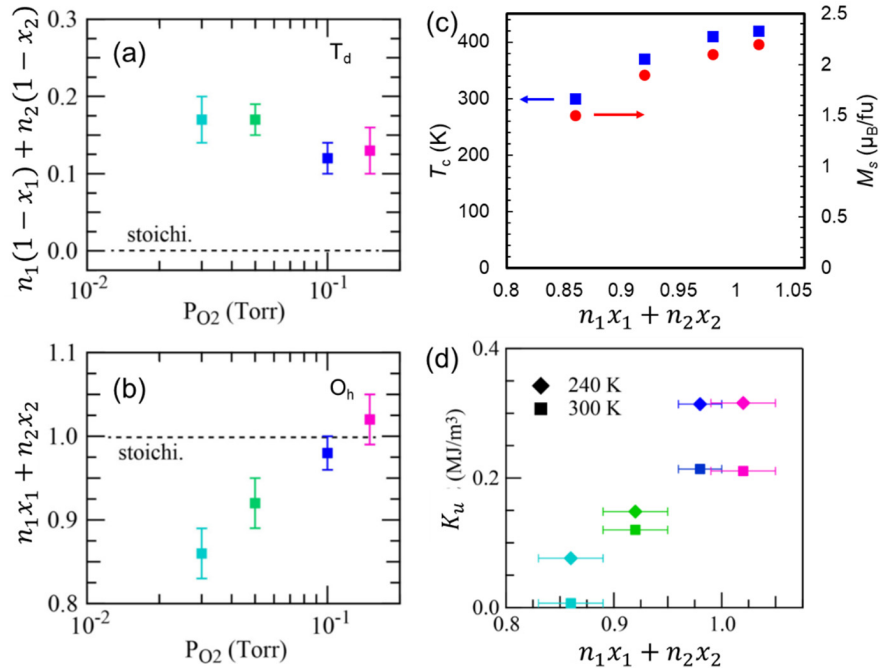

Figure 9. Site Occupancy and magnetic properties of NCO(001)/MAO(001) films (about 30 nm) grown in different $O_2$ pressure. (a)-(b) Occupancy of Ni on the $O_h$ (a) and $T_d$ (b) sites measured using resonance x-ray diffraction. (c) $T_C$ and low-temperature $M_s$ versus Ni occupancy on the $O_h$ site. (d) Uniaxial PMA energy ($K_u$) as a function of Ni occupancy on the $O_h$ site. Adapted from Ref. [47]. Copyright 2020 American Physical Society.

### 3.2.3 Effect of Inhomogeneous Site Occupancy

Besides cation vacancy, inhomogeneous site occupancy has also been observed in NCO films grown in low $O_2$ pressure. More specifically, the $T_d$ site vacancy occurs in a phase-separated fashion, effectively consisting of regions with optimal phase and regions with $T_d$-vacant phase [Fig. 10(a)].[40,48] In the optimal NCO phase, $M_s$ is parallel to the moments on the $T_d$ site and antiparallel to that on the $O_h$ site. This is because the moments on the $O_h$ and $T_d$ sites are antiferromagnetically coupled, and all cations ($Ni^{2+}$, $Ni^{3+}$, $Co^{2+}$, and $Co^{3+}$) have higher spin magnetic moments when they are on the $T_d$ site (Table II). For the $T_d$-vacant phase, the magnetic contribution may be dominated by the $O_h$ sites, and $M_s$ may be aligned with the magnetic moment on the $O_h$ sites.

Given the antiferromagnetic coupling between the moments on the $O_h$ and $T_d$ sites, the magnetization between the neighboring optimal phase and the $T_d$-vacant phase are expected to be antiparallel when there is no external field (Fig. 10(c)). This is observed in the "exchange spring"-type[95,96] magnetic hysteresis loops (Fig. 10(d)).[48] There are clearly two contributions to the hysteresis loop that are antiferromagnetically coupled. When the external field is large enough, both contributions are aligned by the field; when the external field is removed, the two contributions become anti-aligned, manifested as a reduction in magnetization near zero field.

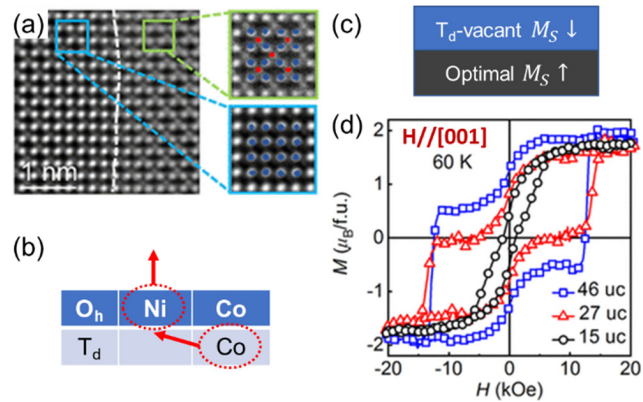

Figure 10. (a) Observation of cation vacancy on the $T_d$ site using scanning transmission electron microscopy for NCO(001)/MAO(001) films (46 unit cell) grown in low $O_2$ pressure. (b) Schematic of the effect of low $O_2$ growth pressure on the site occupancy in NCO. The arrows illustrate Ni leaving the NCO film and Co migrating from the $T_d$ site to the $O_h$ vacant site. (c) Schematic of the antiferromagnetic coupling between the optimal phase and the $T_d$-vacant phase in NCO(001)/MAO(001) films grown in low $O_2$ pressure. (d) Magnetization hysteresis showing "exchange spring" behavior due to the antiferromagnetic coupling illustrated in (c). Adapted from Ref. [48]. Copyright 2020 American Chemical Society.

## 4. Effect of Disorder on the Electronic Properties

Optimal NCO exhibit low resistivity (about 0.8 mΩ cm at 50 K) with weak temperature dependence, making it an appealing choice as transparent conductors.[10,11,18,24,25] The electronic properties of NCO can be sensitively tuned by disorder,[9] which can lead to a metal-insulator transition. Figures 11(a)-(b) show the temperature-dependence of resistivity $\rho(T)$ for NCO films on (001) MAO substrates deposited at different temperatures.[35] The films at high growth temperatures (500°C and higher) exhibit insulating behaviors, while optimal resistivity is observed in the film deposited at 350°C. There is a clear relation between the metallicity of the sample and its structural and magnetic properties, with the most conductive sample possessing the highest $M_s$ and $T_C$.[35,40] Studies have shown that the metallic behavior of NCO is well correlated with the $Ni^{3+}$ concentration (Fig. 11(c)),[22,97] suggesting that the charge itinerancy in NCO originates from the $Ni^{3+}$-$O^{2-}$-$Ni^{3+}$ double exchange.

Besides growth temperature, disorder can also be introduced into epitaxial thin films via strain. Figure 11(d) shows $\rho(T)$ for NCO(111) films deposited on MAO(111) and $Al_2O_3$(001) substrates,[34] which are subject to 0.3% and 4% compressive strain, respectively. The insulating behavior of the film on $Al_2O_3$ shows moderate improvement upon post-growth annealing, ruling out the contributions of valence mixing and spinel inversion. It has been suggested that the possible culprit is the presence of antiphase boundaries upon structural relaxation. This scenario is corroborated by the large magnetoresistance, as will be discussed in Section 6.

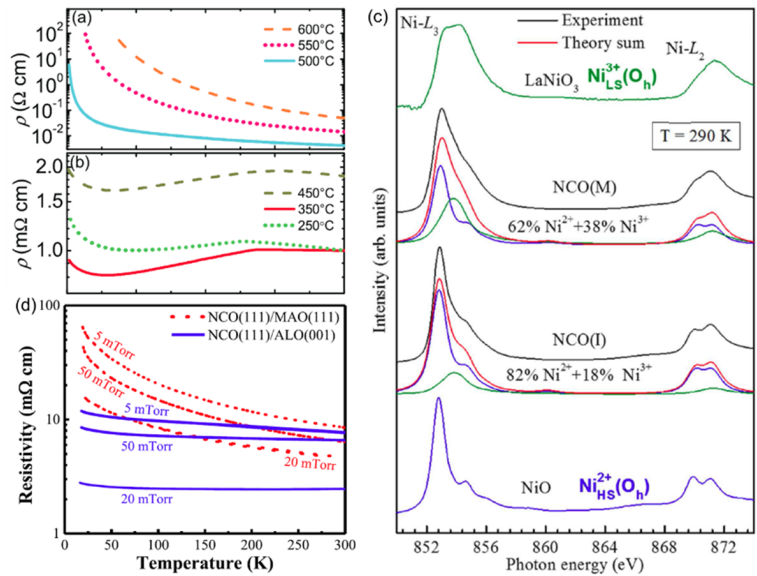

Figure 11. (a)-(b) $\rho$ vs. $T$ for NCO films on MAO(001) substrates grown at different temperatures. Adapted from Ref. [35]. Copyright 2012 AIP Publishing LLC. (c) XMCD spectra of Ni $L_{2,3}$ for metallic (M) and insulating (I) NCO(001) films on MAO substrates. Bital *et al.*, Scientific Reports **5**, 15201 (2015).[22] Copyright 2020 Author(s), licensed under a Creative Commons Attribution (CC BY) license. (d) $\rho$ vs. $T$ for NCO(111) films grown on $Al_2O_3$(001) and MAO(111) substrates at different oxygen growth pressure. Adapted from Ref. [34]. Copyright 2018 IOP Publishing Ltd.

5. Effect of Film Thickness

For epitaxial thin films, film thickness is a powerful control knob for property tuning.[12,37,98] For example, ferroic materials are known to show the finite size effect, *i.e.*, the suppression of ferroic order at the ultrathin limit. Figure 12(a) shows the temperature dependence of sheet resistance $R_\square(T)$ taken on 1.5-30 unit cell thick NCO films deposited on (001) MAO substrates.[12,99] The thicker films (20-30 uc) show optimal conduction, with metallic behavior and $\rho \sim 0.8$ m$\Omega$ cm at 50 K (Fig. 12(b)). A resistance upturn occurs at low temperature, which can be due to weak localization or enhanced charge correlation.[100] The films become progressively more resistive with reduced thickness, transitioning to the strongly localized insulating behavior at 2 uc, with the sheet resistance reaching the two-dimensional quantum resistance of $h/e^2$.[100–102] Similar thickness driven metal-insulator transition has been widely observed in correlated oxides, including (La,Sr)MnO$_3$,[103] $R$NiO$_3$ ($R$: rare earth),[104–106] and SrIrO$_3$.[100,107,108] Such a transition can be due to the surface/interface induced disorder/distortion,[105,106,108,109] dimensionality crossover,[100,104,107,108] or enhanced correlation energy[107] in the ultrathin limit.

The film thickness tuning of charge itinarancy in NCO is accompanied by the modulation of magnetic properties, with $T_C$ and $M_s$ also suppressed in thinner films (Fig. 3(b)).[12,37] On the other hand, a recent study shows for optimal NCO, $T_C$ remains above 300 K in 3 uc thick films, and the magnetic dead layer thickness can be as thin as 1.5 uc (1.2 nm), making NCO highly competitive for spintronic applications compared with other correlated oxides and various two-dimensional van der Waals magnets.[99] The film thickness can also effectively tune the coercive field $H_c$ (Figs. 3(b) and 13(c)). Figure 12(d) shows the switching hysteresis of anomalous Hall resistance at 300 K for 5-30 uc (001) NCO films. While all films exhibit strong PMA with $T_C$ above 300 K, $H_c$ decreases significantly in thinner films, which may be due to the reduced magnetic double well energy as $T_C$ is approaching room temperature in these samples (Fig. 11(c)). Well below $T_C$, the 30 uc film exhibits a $T$-independent $H_c$ of ~2 kOe (Fig. 11(c)). In thiner NCO films, $H_c$ increases exponentially with reducing $T$,[12,99] suggesting thermally activated domain wall depinning.[110] The distinct $T$-dependences of $H_c$ between the thick and thin films can be attributed to enhanced contribution from film surface, where domain nucleation and domain wall pinning can be caused by surface roughness, reconstruction, and spin fluctuation.

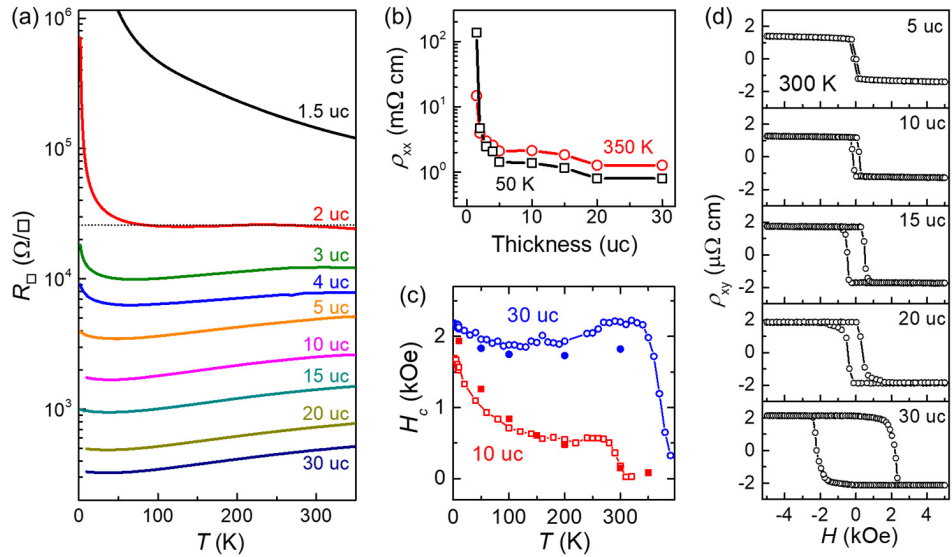

Figure 12. Transport and magnetic properties of optimal (001) NCO films on MAO(001) substrates. (a) Sheet resistance vs. $T$ for 1.5-30 uc films. The dotted line marks $h/e^2$. (b) $\rho_{xx}$ at 50 K and 350 K vs. film thickness. Data in (a) and (b) are taken from Refs. [12,99]. (c) $H_c$ vs. $T$ for 10 and 30 uc films extracted from $M(H)$ (solid symbols) and AHE data (open symbols). (d) $\rho_{xy}$ vs. $H$ at 300 K for 5-30 uc films. (c)-(d): Adapted from Ref. [12]. Copyright 2019 John Wiley & Sons, Inc.

## 6. Magnetotransport

For magnetic conductors, the magnetotransport properties can reveal critical information about the electronic band and spin scattering mechanisms. In this section, we discuss the intringing magnetotransport properties observed in epitaxial NCO films, including linear magnetoresistance and film thickness-/temperature-driven sign change in anomalous Hall effect (AHE). These highly tunable phenomena reflect the complex interplay between band intrinsic Berry phase, SOC, disorder induced spin scattering, and correlation effect.

### 6.1 Magnetoresistance

The magnetoresistance (MR) of NCO films is highly sensitive to disorder and can exhibit distinct magnetic field dependences. With optimal quality, NCO films exhibits very small negative MR in a perpendicular magnetic field with a quasi-linear field dependence. For example, it has been shown that the MR ratio (MRR) of high quality NCO films exhibit very weak $T$-dependence, with the out-of-plane MRR remaining less than 1% at 5 T over a wide temperature range (300 mK-300 K).[12] Figures 13(a)-(b) show the MR taken on 10 uc NCO films at two different temperatures. Switching hysteresis is observed in out-of-plane MR upon magnetization switching. Above $H_c$, the sample can sustain the linear MR up to 17 T (Fig. 13(b)). In contrast, the in-plane MR shows a typical parabolic dependence at low magnetic field (Fig. 13(a)), which can be attributed to magnetization rotation to a hard axis. It has been suggested that the linear MR shares similar origins as the intrinsic contribution to the AHE,[99] which originates from the band-intrinsic Berry curvature in systems with broken inversion system. It is favored in magnetic conductors with large SOC and can persist to higher magnetic fields in more disordered materials. It is also noted that at ultra-low temperature, there are sharp spikes appearing in MR at low magnetic field (Fig. 13(b)), which may signal the emergence of unusual spin texture in the presence of strong SOC.

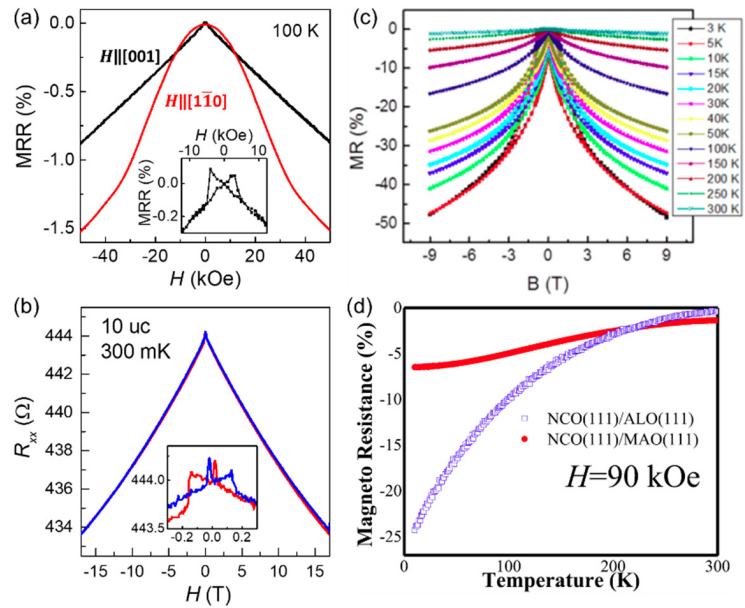

Figure 13. (a)-(b) MR for 10 uc optimal (001) NCO films on MAO(001) substrates. (a) MRR vs. $H$ at 100 K in in-plane [1$\bar{1}$0] and out-of-plane [001] magnetic fields. Adapted from Ref. [12]. Copyright 2019 John Wiley & Sons, Inc. (b) $R_{xx}$ vs. $H$ at 300 mK in out-of-plane magnetic fields.

(c) In-plane MRR vs. *H* for an 18 nm NCO(001) film on MgO(001) substrate at various temperatures. Adapted from Ref. [40]. Copyright 2017 American Chemical Society. (d) MR vs. *T* for NCO(111) films on MAO(111) and $Al_2O_3$(111) substrates. Adapted from Ref. [34]. Copyright 2018 IOP Publishing Ltd.

For highly disordered NCO films, on the other hand, the MR is significantly enhanced and exhibits strong temperature dependence (Figs. 14(c)-(d)),[34,40,94,111] and the magnitude and temperature-dependence of MR strongly resemble those of nano-crystalline NCO samples.[112] It has been attributed to spin-polarized transport through antiphase boundaries[34] or spin filtering effect in phase separated samples[40,48], with the latter similar to that of the colossal magnetoresistive manganites.[113] The magnitude of MR can thus be effectively tuned by the level of disorder via changing the growth condition or substrate strain (Fig. 13(d)).[34,40,111]

## 6.2 Anomalous Hall Effect

Epitaxial NCO films also exhibit intriguing AHE that can be tuned by film thickness,[12,99,114] as well as topological Hall-like features.[98] Figure 14(a) shows the Hall resistance $R_{xy}$ in out-of-plane magnetic field taken on the 30 uc NCO film shown in Fig. 3(b), which exhibits clockwise switching hysteresis of AHE.[12] The anomalous Hall resistance $R_{AHE}$ shows a non-monotonic temperature-dependence, peaking close to room temperature. For a 10 uc NCO film, $R_{AHE}$ also peaks at ~300 K and gradually diminishes with reducing temperature, reaching zero at about 190 K (Fig. 14(b)).[12] When the temperature is further lowered, the AHE re-emerges with counter-clockwise hysteresis, corresponding to a sign change in $R_{AHE}$ from negative to positive. Such temperature-driven sign change does not correspond to any abrupt change in the magnetic state (Fig. 3(b)) or the change of carrier-type. Unlike the sign change phenomena observed in SrRuO3,[115,116] there is no universal scaling relation between the anomalous Hall conductivity $\sigma_{xy}$ and magnetization,[12,114] ruling out the impact of sign reversal in the Berry curvature. It has thus been proposed that the sign change is due to the competing effects of different AHE mechanisms, which possess different signs and temperature dependences.[12,117,118]

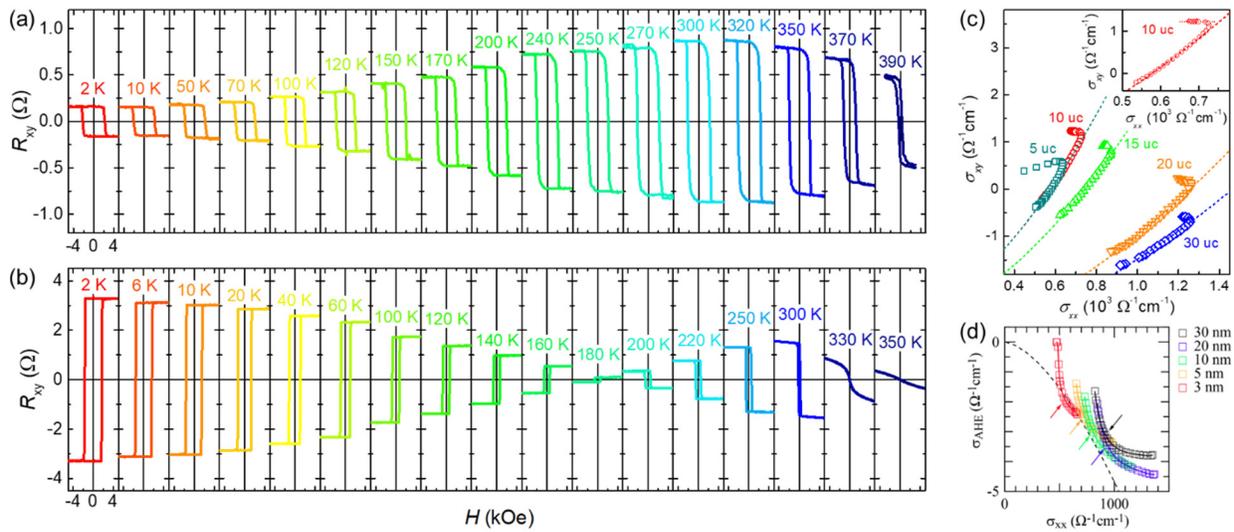

Figure 14. AHE in (001) NCO films on (100) MAO substrates. (a)-(b) $R_{xy}$ vs. *H* for (a) 10 uc and (b) 30 uc films at various temperatures. (c) $\sigma_{xy}$ vs. $\sigma_{xx}$ for 5-30 uc NCO films. The dashed lines

correspond to Eq. (6). Adapted from Ref. [12]. Copyright 2019 John Wiley & Sons, Inc. (d) $\sigma_{AHE}$ vs. $\sigma_{xx}$ for 3-30 nm NCO films. The dashed line corresponds to $\sigma_{AHE} \sim \sigma_{xx}^{1.6}$. Adapted from Ref. [114]. Copyright 2021 American Physical Society.

Previous theoretical studies have shown that for multi-band magnetic conductors, the scaling behavior of AHE can be divided into three regimes depending on how clean the system is.[119,120] For dirty metals, with the relaxation rate $\hbar/\tau$ exceeding Fermi energy $E_F$, $\sigma_{xy}$ is dominated by impurity spin scattering and exhibits a power-law dependence on the longitudinal conductivity $\sigma_{xx}$: $\sigma_{xy} \sim \sigma_{xx}^{1.6}$.[121] For moderately dirty systems ($\hbar/\tau \leq E_F$), $\sigma_{xy}$ is dominated by the band intrinsic Berry phase effect and becomes insensitive to scattering, remaining a constant ($\sigma_{xx}^{(0)}$). For super-clean systems, skew scattering dominates and $\sigma_{xy}$ scales linearly with $\sigma_{xx}$. As the conductivity of optimal NCO films is close to the boundary between the dirty metal and moderately dirty regimes ($\sigma_{xx} \sim 10^3 \ \Omega^{-1} \text{cm}^{-1}$), it has been suggested that both impurity scattering and Berry phase effect can contribute to the AHE. The relative strength of these two can be tuned by temperature and film thickness (Figs. 14(a)-(b)), leading to the sign change. As shown in Fig. 14(c), the $\sigma_{xy}$ vs. $\sigma_{xx}$ scaling relation for 5-30 uc NCO films can be well described by:

$$\sigma_{xy} = \sigma_{xy}^{(0)} + A\sigma_{xx}^{1.6}. \tag{6}$$

Here the first and second terms depict the Berry phase and impurity scattering contributions, respectively, and $A$ is a fitting parameter that increases with reducing film thickness.[12] It has been further showed that Eq. (6) is sustained down to the 2 uc films in the metallic phase, while the Berry phase contribution vanishes as the samples transitions to the insulating phase.[99] In contrast, for samples with $\sigma_{xx}$ well above $10^3 \ \Omega^{-1} \text{cm}^{-1}$, the Berry phase contribution dominates $\sigma_{xy}$ (Fig. 14(d)).[114] The clear correlation between the metallicity of sample and the relative strength between the first and second terms in Eq. (6) yields strong support to this scenario. Another interesting observation is the emergence of another scattering independent $\sigma_{xy}$ regime at low temperature (Fig. 14(c)), where the conduction is dominated by disorder enhanced correlation effect and the side-jump mechanism may also contribute to AHE.[122–124]

## 7. Conclusion and Future Perspective

NCO is a versatile material with robust magnetic order and close to half-metallicity. It hosts a plethora of cation configurations in terms of local oxygen environments ($O_h$ and $T_d$), cation species (Ni and Co), and valence states (2+ and 3+). The small energy difference between these configurations leads to high electrochemical activity, making NCO appealing for catalysis and energy storage applications.[11] While the inverse spinel structure is unstable in the large single crystal form, there has been rapid development of high quality epitaxial thin films in recent years leveraging the strong spin-lattice coupling. Its electronic and magnetic properties depend sensitively on the strain and disorder levels — substrate type, growth condition, and film thickness have been exploited to effectively control the metallicity, magnetic transition temperature, magnetization, coercive field, magnetic anisotropy, and magnetotransport of the samples. The highly tunable nature yields NCO distinct advantages for manipulating spin using non-magnetic means, which is highly desirable for developing energy efficient spintronic applications.

One promising application for NCO is to serve as the spin-injection layer for epitaxial magnetic tunnel junctions (MTJ). It has been predicted theoretically that spinel MTJs with MAO tunnel barriers can facilitate spin-filtering and enhance tunnel magnetoresistance.[125,126] Studies of epitaxial MTJs composed of MAO and NCO also reveal high spin polarization of -73%.[13] The above room temperature $T_C$, strain tunable PMA,[15] film thickness dependent $H_c$, close to half metallicity, and fast spin dynamics[14] make NCO films ideal for constructing high efficiency, high speed MTJs.

Another possible device concept for NCO is magnetic field sensors[127]. Optimal NCO exhibits linear MR with weak temperature dependence that can persist to high magnetic field (Fig. 13(a)-(b)).[12,48] The magnitude of MR can be further tuned by disorder and strain,[34,40] which can be exploited to design field sensing devices with high sensitivity.

Epitaxial NCO films also present a rich playground to explore the AHE.[128] The conductivity of optimal NCO is close to the boundary between the dirty metal and moderately dirty regimes, where multiple mechanisms can contribute to AHE.[119,120] As the electronic properties can be senstiviely tuned by the disorder level via changing substrate type, growth condition, and film thickness, the relative strength between the contributions from the band intrinsic Berry phase effect, impurity scattering, and correlation energy can be varied. The anomalous Hall conductivity of NCO is comparable with that of magnetic semiconductors,[129] while it can persist well above room temperature. These verstile features make NCO an appealing material choice for exploring AHE-based device applications.

As an emerging spintronic material, there are three major directions for future studies of NCO. The first one involves further improving the material quality, particularly magnetic properties. It is of both fundamental and technologogical interests to explore the finite size limit of magnetism in NCO. To date, the thinnest NCO films with $T_C > 300$ K is 3 uc (2.4 nm).[99] Robust PMA has been observed in NCO films as thin as 1.5 uc (1.2 nm), where $T_C$ is suppressed to about 170 K.[99] The thickness scaling behavior of these ultrathin NCO films outperform the widely investigated magnetic oxides such as $SrRuO_3$[130] and $(La,Sr)MnO_3$[103] and are comparable to the two-dimensional van der Waals magnets with PMA.[131] The scalable growth method via physical vapor deposition make them highly competitive for developing high density magnetic memories. It is thus important to explore material strategies to enhance $T_C$ and engineer the magnetic parameters in the ultrathin NCO films.

The second direction centers on gaining microscopic understanding of the unusual properties of these materials. High qulaity epitaxial NCO thin films only become available over the last ten years. Due to its relatively large unit cell, theoretical computation of band properties of spinels is lagging behind the perovskite oxides.[39] For example, the emergence of unconventional low field spikes in MR (Fig. 13(b)) and possible topological Hall signal[111] suggest the existence of non-coplanar spin textures, which may be associated with non-trivial band topology. To elucidate the origin of these effects, it requires combined theoretical and experimental efforts to map out the band structure, determine the strength of SOC and Dzyaloshinskii-Moriya interaction, and identify the possible existence of chiral spin textures. It has also been shown that the thermoelectric coefficents of NCO films is much smaller than other spinel oxides,[132] while whether it is intrinsic to NCO or compromised by the sample quality remains to be investigated.

The third direction lies on developing novel device concepts, exploiting the magnetotransport anomaly and strain tunable properties in NCO films. The fact that the magnitude and size of the anomalous Hall resistance as well as the switching field can be sensitively tuned opens the door for developing novel information storage devices using AHE. It is desirable to explore possible electric field effecct control of AHE, which can be built on the field effect transistor device architechture.[133,134] The field effect can also be exploited to tune the topological Hall effect and, if confirmed, chiral spin textures, which can be of interest for topological computing. As the PMA in NCO can be effectively tuned by strain, it is also conceivable to interface NCO-based MTJ with a piezoelectric substrate, exploring electric control of the switching characteristics of the tunneling magnetoresistance.[67]

In conclusion, in this Perspetive, we summarize the recent advacements on understanding the intriguing properties of epitaxial NCO films, discussing both the optimal configuration and the strong tunability. These understandings allow us to outline a few promising directions of research, aiming at inspiring future studies of this emerging material system for high density, high speed, and energy efficient spintronic and nanoelectronic applications.

**Acknowledgement:**

The authors acknowledge the primarily support by National Science Foundation (NSF) through EPSCoR RII Track-1: Emergent Quantum Materials and Technologies (EQUATE), Award OIA-2044049. Additional supports were from NSF Grant No. DMR-1454618 and Grant No. DMR-1710461. Work by Z. G. C. was supported by National Key R&D Program of China Grant No. 2018YFA0305604 and National Science Foundation of China (NSFC) No. 11874403.

Rev. B **84**, 165304 (2011).

[122] I.A. Ado, I.A. Dmitriev, P.M. Ostrovsky, and M. Titov, "Sensitivity of the anomalous Hall effect to disorder correlations," Phys. Rev. B **96**, 235148 (2017).

[123] A.A. Kovalev, J. Sinova, and Y. Tserkovnyak, "Anomalous Hall Effect in Disordered Multiband Metals," Phys. Rev. Lett. **105**, 36601 (2010).

[124] T. Fukumura, H. Toyosaki, K. Ueno, M. Nakano, T. Yamasaki, and M. Kawasaki, "A Scaling Relation of Anomalous Hall Effect in Ferromagnetic Semiconductors and Metals," Jpn. J. Appl. Phys. **46**, L642 (2007).

[125] H. Sukegawa, Y. Miura, S. Muramoto, S. Mitani, T. Niizeki, T. Ohkubo, K. Abe, M. Shirai, K. Inomata, and K. Hono, "Enhanced tunnel magnetoresistance in a spinel oxide barrier with cation-site disorder," Phys. Rev. B **86**, 184401 (2012).

[126] J. Zhang, X.-G. Zhang, and X.F. Han, "Spinel oxides: Δ1 spin-filter barrier for a class of magnetic tunnel junctions," Appl. Phys. Lett. **100**, 222401 (2012).

[127] I. Žutić, J. Fabian, and S. Das Sarma, "Spintronics: Fundamentals and applications," Rev. Mod. Phys. **76**, 323 (2004).

[128] N. Nagaosa, J. Sinova, S. Onoda, A.H. MacDonald, and N.P. Ong, "Anomalous Hall effect," Rev. Mod. Phys. **82**, 1539 (2010).

[129] T. Jungwirth, Q. Niu, and A.H. MacDonald, "Anomalous Hall Effect in Ferromagnetic Semiconductors," Phys. Rev. Lett. **88**, 207208 (2002).

[130] J. Xia, W. Siemons, G. Koster, M.R. Beasley, and A. Kapitulnik, "Critical thickness for itinerant ferromagnetism in ultrathin films of SrRuO3," Phys. Rev. B **79**, 140407 (2009).

[131] Z. Fei, B. Huang, P. Malinowski, W. Wang, T. Song, J. Sanchez, W. Yao, D. Xiao, X. Zhu, A.F. May, W. Wu, D.H. Cobden, J.-H. Chu, and X. Xu, "Two-dimensional itinerant ferromagnetism in atomically thin Fe3GeTe2," Nat. Mater. **17**, 778 (2018).

[132] H. Koizumi, A. Hidaka, T. Komine, and H. Yanagihara, "Anomalous nernst and seebeck effects in nico2 o4 films," J. Magn. Soc. Japan **45**, 37 (2021).

[133] H. Mizuno, K.T. Yamada, D. Kan, T. Moriyama, Y. Shimakawa, and T. Ono, "Electric-field-induced modulation of the anomalous Hall effect in a heterostructured itinerant ferromagnet SrRuO3," Phys. Rev. B **96**, 214422 (2017).

[134] D. Chiba, A. Werpachowska, M. Endo, Y. Nishitani, F. Matsukura, T. Dietl, and H. Ohno, "Anomalous Hall Effect in Field-Effect Structures of (Ga,Mn)As," Phys. Rev. Lett. **104**, 106601 (2010).
31